\documentstyle[amssymb,aps,preprint]{revtex}
\begin{document}
\title{Phase separation and valence instabilities in cuprate superconductors.
Effective one-band model approach.}
\author{M. E. Sim\'{o}n and A. A. Aligia}
\address{Centro At\'{o}mico Bariloche and Instituto Balseiro \\
Comisi\'on Nacional de Energ\'{\i}a At\'{o}mica \\
8400 Bariloche, Argentina }
\maketitle

\begin{abstract}
We study the Cu-O valence instability (VI) and the related phase separation
(PS) driven by Cu-O nearest-neighbor repulsion $U_{pd}$, using an effective
extended one-band Hubbard model ($H_{eff}$) obtained from the extended three-band
Hubbard model, through an appropriate low-energy reduction. $H_{eff}$ is solved by
exact diagonalization of a square cluster with 10 unit cells and also within
a slave-boson mean-field theory. Its parameters depend on doping for 
$U_{pd}\neq 0$ or on-site O repulsion $U_p\neq 0$. The results using both
techniques coincide in that there is neither VI nor PS for doping
levels $x<0.5$ if $U_{pd}\lesssim 2$ eV. The PS region begins for 
$U_{pd}\gtrsim 2$ eV at large doping $x>0.6$ and increases with increasing $%
U_{pd}$. The PS also increases with increasing on-site Cu repulsion $U_d$.

\end{abstract}

\newpage

\section{Introduction}

It is widely accepted that the electronic properties of the cuprates are
well described by the three-band Hubbard model $H_{3b}$ \cite{eme,var,little}%
. Zhang and Rice \cite{zha} suggested that this model can be reduced to an
effective one-band $t-J$ model under certain simplifying hypothesis: small
O-Cu hopping $t_{pd},$ and zero on-site O Coulomb repulsion $U_p$, Cu-O
Coulomb nearest-neighbor interatomic repulsion $U_{pd}$ and O-O hopping $%
t_{pp}$. While the one-band effective models can explain the magnetic
properties of high-$T_c$ materials and can provide magnetic mechanisms for
superconductivity \cite{rieri,rvb}, these models, with fixed parameters, do
not take into account the effects of the Cu-O repulsion $U_{pd}$. The
density of carriers in cuprate superconductors is very low. The average
distance between two holes is larger than 7 \AA\ and the Cu-O distance is $%
\sim $ 1.9 \AA\ and therefore $U_{pd}$ is expected to be poorly screened.
Neglecting screening $U_{pd}\sim 7$ eV, while direct calculations give $%
U_{pd}\sim 3$ eV \cite{rush} and constrained-density-functional results
predict $U_{pd}\sim 1$ eV \cite{lda}. Thus, as pointed out early after the
discovery of high-$T_c$ superconductivity, the effects of $U_{pd}$ can be
important \cite{var,cluster}. These effects include a charge-transfer
mechanism for superconductivity \cite{little,cluster,sudbo}, and marginal
-Fermi-liquid behavior \cite{perakis}. It is known that a sufficiently large 
$U_{pd}$ induces a charge-transfer instability (CTI) and a valence
instability (VI) related with Cu-O charge transfer \cite{hicks}. These
instabilities are associated with phase separation because of the coupling
between the fluctuations of the valence and the total density \cite
{bang,raimondi}.

The three-band Hubbard model including $U_{pd}$ has been studied by exact
diagonalization of small clusters \cite{cluster}, random-phase approximation 
\cite{little,bang,little2,kothe}, Gutzwiller variational wave function \cite
{copper}, and slave-bosons with 1/N expansion, for the particular case $%
U_d=\infty $ and with $U_p$ and $U_{pd}$ treated in the Hartree
approximation \cite{raimondi,hira,capra}. The results of these works are
qualitatively similar in general. Increasing $U_{pd}$ the system reaches the
CTI, VI and the related phase separation. Near the phase-separation
boundary, the effective interaction in the Cooper {\it s} and {\it d}
channels becomes attractive.

Bang {\it et al.} in weak coupling \cite{bang} and Raimondi {\it et al}. in
strong coupling \cite{raimondi} have shown that even though the energy $%
\omega _{exc}$ of the charge-transfer collective excitonic mode at zero wave
vector, decreases with increasing $U_{pd}$, it remains finite at the CTI.
However, this mode is coupled to the zero-sound mode, leading to a
charge-transfer-mode mediated attraction which allows the violation of the
Landau stability criterion $F_0^S>-1$. This leads to the simultaneous
divergence of the compressibility and charge-transfer static susceptibility
indicating that phase separation and the CTI take place.

In spite of the qualitative agreement on the facts mentioned above among the
different techniques, the regions of phase separation differ in the strong
and weak coupling approaches, and the critical value of $U_{pd}$ where the
instabilities take place depends strongly on the approximation used. Also,
the effect of the variation of the parameters has not been investigated in
detail and realistic values of $U_d$ have not been studied so far. Thus,
further research on these subjects seems necessary.

On the other hand, it is important to address the question of to what extent
the above mentioned properties of the three-band Hubbard model can be
accounted for using a one-band effective model derived from the former through
a sufficiently accurate mapping procedure. Generalized $t-J$ and one-band
Hubbard models have been obtained performing systematic low-energy
reductions of the three-band Hubbard \cite{fed,jef,fei,hay,sim1,sim2,bel,ali}%
. These reductions either are based on a cell  method \cite
{fed,jef} or use an effective spin-fermion model with renormalized
parameters as an intermediate step in the derivation \cite{ali,bati}. In
contrast to the original derivation of Zhang and Rice, these methods allow
to extend the mapping procedure to realistic and large values of Cu-O
hopping $t_{pd}$, and show that the Zhang-Rice singlets are stable for large
Cu-O covalency. There have been an important amount of research devoted to
the study of the validity of the low-energy reduction and the stability of
the Zhang-Rice singlets. In particular, the roles of local triplets states 
\cite{tripletes} and apical O ions \cite{api} have been recently
investigated. Exact diagonalization of small clusters have shown that the
low-energy spectrum \cite{tripletes,bati1} and magnetic properties \cite{ero}
of the three-band model can be well reproduced by a $t-t^{\prime }-t^{\prime
\prime }-J$ model, where $t^{\prime }$ is the next-nearest-neighbor hopping
and $t^{\prime \prime }$ is a next-nearest-neighbor hopping combined with a
nearest-neighbor spin flip. This three-body term, with the sign
corresponding to large O-O hopping $t_{pp}$ in the original three-band
model, can stabilize a resonance-valence-bond superconducting ground state 
\cite{rvb}.

In this paper, starting from the three-band Hubbard model, we derive an
effective generalized one-band Hubbard model , using the cell
method generalized to take into account properly the intercell part of the O
intratomic repulsion $U_p$ and the Cu-O interatomic repulsion $U_{pd}$.
As a consequence of these interactions, the parameters of the effective
model become dependent on the particle density. Solving the effective
one-band model exactly in a $\sqrt{10}\times \sqrt{10}$ cluster, and also in
the slave-boson mean-field theory of Kotliar and Ruckenstein \cite{kot} we
study the above mentioned valence instability (VI), the regions of phase
separation (PS), and also the dependence of the effective one-band
parameters with $U_{pd}$ and doping. The advantage of our approach in
comparison with other methods discussed above for the study of the VI, CTI
and PS, is that the intracell correlations, including Cu on-site
repulsion $U_d$ and a large part of the $U_{pd}$ and $U_p$ terms, are taken
into account exactly at each cell. This allows us in particular to take
realistic values of $U_d$ ($\sim 10$ eV), while in previous treatments $U_d$ was
either small or infinite.

In section II we describe the different Hamiltonians and the method for
low-energy reduction. Section III contains the results and section IV is 
a short summary and discussion.

\section{Reduction from the three-band to a one-band Hamiltonian}

We start from the three-band extended Hubbard model: 
\begin{eqnarray}
H_{3b} &=&\Delta \sum_jp_{j\sigma }^{\dagger }p_{j\sigma
}+U_d\sum_id_{i\uparrow }^{\dagger }d_{i\uparrow }d_{i\downarrow }^{\dagger
}d_{i\downarrow }+U_p\sum_jp_{j\uparrow }^{\dagger }p_{j\uparrow
}p_{j\downarrow }^{\dagger }p_{j\downarrow }  \nonumber \\
&&+U_{pd}\sum_{i\delta \sigma \sigma ^{\prime }}d_{i\sigma ^{\prime
}}^{\dagger }d_{i\sigma ^{\prime }}p_{i+\delta \sigma }^{\dagger
}p_{i+\delta \sigma }\ +t_{pd}\sum_{i\delta \sigma }(p_{i+\delta \sigma
}^{\dagger }d_{i\sigma }+h.c.)-t_{pp}\sum_{j\gamma \sigma }p_{j+\gamma
\sigma }^{\dagger }p_{j\sigma }\ 
\end{eqnarray}
The sum over $i$ ($j$) runs over all Cu (O) ions. The vector $\delta $ ($%
\gamma $) connects a Cu (O) site with one of its four nearest O atoms. The
operator $d_{i\sigma }^{\dagger }$ ($p_{j\sigma }^{\dagger }$) creates a
hole with symmetry $d_{x^2-y^2}$ ($p_\sigma $) at site $i$ ($j$) with spin $%
\sigma $. The phases of half of the orbitals have been changed in such a way
that for all directions, the hoppings are positive ($t_{pd}$, $t_{pp}>0$).
The parameters of the model are known approximately from
constrained-density-functional calculations \cite{lda}.

The first step in the cell method \cite{jef,sim1,bel} is to
change the basis of the O orbitals to linear combinations which hybridize ($%
\alpha _{k\sigma }$ ) and do not hybridize ($\gamma _{k\sigma }$ ) with $%
d_{k\sigma }$ orbitals, due to the term in $t_{pd}$ at each point {\bf k} of
the reciprocal space. We denote the Wannier functions of these orbitals as $%
\alpha _{i\sigma }$ and $\gamma _{i\sigma }$ respectively. The $\gamma
_{i\sigma }$ ({\it non-bonding} states) lie very high in energy and are
neglected. After the change of basis, the original Hamiltonian can be
separated in two parts, one containing the intracell (and generally larger)
terms and another containing the intercell terms: 
\begin{equation}
H_{3b}=\sum_iH_i+H_{inter},  \label{3b}
\end{equation}
with 
\begin{eqnarray}
H_i &=&(\Delta -\mu (0)t_{pp})\sum_\sigma \alpha _{i\sigma }^{\dagger
}\alpha _{i\sigma }\   \nonumber \\
&&+U_dd_{i\uparrow }^{\dagger }d_{i\uparrow }d_{i\downarrow }^{\dagger
}d_{i\downarrow }+2t_{pd}\lambda (0)\sum_\sigma (d_{i\sigma }^{\dagger
}\alpha _{i\sigma }+h.c.)  \nonumber \\
&&+U_{pd}f(0)\sum_{\sigma \sigma ^{\prime }}d_{i\sigma ^{\prime }}^{\dagger
}d_{i\sigma ^{\prime }}\alpha _{i\sigma }^{\dagger }\alpha _{i\sigma
}+U_ph(0)\alpha _{i\uparrow }^{\dagger }\alpha _{i\uparrow }\alpha
_{i\downarrow }^{\dagger }\alpha _{i\downarrow }  \label{cell} \\
H_{inter} &=&\sum_{i\neq j\sigma }[2t_{pd}\lambda (j-i)d_{j\sigma }^{\dagger
}\alpha _{i\sigma }-t_{pp}\mu (j-i)\alpha _{j\sigma }^{\dagger }\alpha
_{i\sigma }+h.c]  \nonumber \\
&&+U_{pd}\sum_{i\neq jl\sigma \sigma ^{\prime }}[f(i-j,i-l)n_{i\sigma
^{\prime }}^d\alpha _{j\sigma }^{\dagger }\alpha _{l\sigma }+h.c.]  \nonumber
\\
&&+U_p\sum_{ijlm}^{\prime }h(i-j,i-l,i-m)\alpha _{i\uparrow }^{\dagger
}\alpha _{m\uparrow }\alpha _{j\downarrow }^{\dagger }\alpha _{l\downarrow }
\label{inter}
\end{eqnarray}
The functions of the lattice vectors $\lambda $, $\mu $, $\nu $, $f$ and $h$
are given in Ref. \cite{bel}. They decay rapidly with increasing argument and
as a consequence, most of the original hoppings and interactions are
contained in $\sum_iH_i$. The $\sum^{\prime }$ indicates that in the last
sum the term with $i=j=m=l$ is excluded.

The ordinary cell method consists in solving $H_i$ exactly in
the subspaces of 0, 1 and 2 particles and retaining the ground state in each
subspace. The non trivial retained eigenstates have the form:

\begin{equation}
|i2\rangle =\frac{A_1}{\sqrt{2}}(d_{i\uparrow }^{\dagger }\alpha
_{i\downarrow }^{\dagger }-d_{i\downarrow }^{\dagger }\alpha _{i\uparrow
}^{\dagger })-A_2\alpha _{i\uparrow }^{\dagger }\alpha _{i\downarrow
}^{\dagger }-A_3d_{i\uparrow }^{\dagger }d_{i\downarrow }^{\dagger
}|0\rangle .  \label{para1}
\end{equation}
\begin{equation}
|i\sigma \rangle =(B_1\ d_{i\sigma }^{\dagger }-B_2\alpha _{i\sigma
}^{\dagger })|0\rangle .  \label{para2}
\end{equation}
The energies of these states will be denoted $\ E_2$ and $E_1$ respectively.
The first term of $|i2\rangle $ corresponds to the Zhang-Rice singlet in the
original derivation \cite{zha}. Our modification consists in considering the
coefficients as variational parameters which are determined minimizing the
total energy. We treat the terms in $U_{pd}$ of $H_{inter}$ with $i\neq
j\neq l$ replacing $n_{i\sigma ^{\prime }}^d$ by its expectation value.
Using the fact that $\sum_if(i-j,i-l)=0$, this is equivalent to the treatment
of Ref. \cite{fei}. Similarly in the terms in $U_p$ of $H_{inter}$ with $i=m$
and $j\neq l$ we replace $\alpha _{i\sigma }^{\dagger }\alpha _{i\sigma }$
by its expectation value. The same treatment is done for $i\neq m$ and $j=l$%
. The remaining terms except those with $i=m$ and $j=l$ were neglected. The
intercell Hamiltonian is written as a function of the retained eigenstates
of $H_i$, and mapping the latter into those of the one-band Hubbard model at
site $i$, one obtains the following effective model: 
\begin{eqnarray}
H\  &=&E_1\sum_{i\sigma }n_{i\sigma }+U\sum_i\ n_{i\uparrow }n_{i\downarrow
}+\sum_{<ij>\sigma }(c_{j\sigma }^{\dagger }c_{i\sigma }\left\{
t_{AA}(1-n_{i,-\sigma })(1-n_{j,-\sigma })\right.   \nonumber \\
&&+t_{AB}[n_{i,-\sigma }(1-n_{j,-\sigma })+n_{i+l,-\sigma }(1-n_{i,-\sigma
})]+t_{BB}n_{i,-\sigma }n_{j,-\sigma }+h.c.  \nonumber \\
&&\sum_{<ij>\sigma \sigma ^{\prime }}V_{11}(1-n_{i,-\sigma })(1-n_{j,-\sigma
^{\prime }})n_{i\sigma }n_{j\sigma ^{\prime }}  \nonumber \\
&&\sum_{<ij>\sigma \sigma ^{\prime }}V_{12}((1-n_{i,-\sigma })n_{j,-\sigma
^{\prime }}+n_{i,-\sigma }(1-n_{j,-\sigma ^{\prime }}))n_{i\sigma
}n_{j\sigma ^{\prime }}  \nonumber \\
&&\sum_{<ij>\sigma \sigma ^{\prime }}V_{22}n_{i,-\sigma }n_{j,-\sigma
^{\prime }}n_{i\sigma }n_{j\sigma ^{\prime }}  \label{1b}
\end{eqnarray}

We obtain the ground state energy of $H$ as a function of the three free
variational parameters (Eqs. \ref{para1} and \ref{para2} ) using two
different approaches:

\begin{enumerate}
\item  The slave-boson approximation (SB) of Kotliar and Ruckenstein \cite
{sim2,kot} after treating the nearest-neighbor repulsions in the
Hartree-Fock approximation.

\item  Exact diagonalization (ED) of a square cluster with 10 unit cells.
\end{enumerate}

\section{Results}

We begin this section by studying the parameters of the one-band effective
model and its dependence with the original parameters and doping. As a basis
for our study we take the parameters of the original three-band Hubbard
model determined from constrained-density-functional approximation \cite{lda}.
We take the unit of energy as $t_{pd}=1\simeq 1.3$ eV. However we also
study the effect of $U_{pd}$ and $U_d$ within a wider range. In the
(unrealistic) limit $U_p$ $=U_{pd}=0,$ the intercell interactions vanish ($%
V_{11}=V_{12}=V_{22}=0$) and the variational parameters $A_i$, $B_i$ (Eqs. 
\ref{para1} and \ref{para2} ) which minimize the total energy correspond to
the eigenstates of the cell Hamiltonian $H_i$ (Eq. \ref{cell}) for all
dopings. Thus the effective one-band parameters are independent of doping.
Increasing $U_{pd}$ and $U_p$, the intercell interactions appear and the
variational parameters start to differ from those corresponding to the
low-energy eigenstates of $H_i$ to take into account better the intercell
interactions. The variational parameters as well as the one-band effective
parameters become doping dependent, although this dependence is very weak
for small $U_{pd}$ and $U_p.$

In Figs. 1-4 we show the one-band parameters as functions of doping for
different values of $U_{pd}$ using ED. The effective on-site repulsion $U$
increases with $U_{pd}$ except for small values of $\Delta $ and high doping
where the amount of O states in the local singlet (Eq. \ref{para1})
increases and the mixed Cu-O part (which pays $\sim U_{pd}$) decreases.
Related with this fact, $U$ decreases with doping for small values of $%
\Delta .$ This reflects a {\it metallization} of the system with a larger
amount of Cu-O covalency which should be reflected in an increase of the
conductivity. This might be related to the sudden increase in the Hall
conductivity observed in La$_{2-x}$Sr$_x$CuO$_4$ for $x>0.17$  \cite{hall}.
As a consequence of the larger amount of O holes in the doubly occupied
cells, and the change in the variational parameters $B_i$ (Eq. \ref{para2})
to minimize the loss of energy due to the Cu-O repulsion $U_{pd},$ the
energy $E_1$ and the amount of O holes in the singly-occupied cells increase with
doping for small $\Delta $ and large $U_{pd}$. For others values of $\Delta $
and $U_{pd}$, $E_1$ is rather insensitive to doping (see Fig. 2).

For small values of $U_{pd}$ the three correlated hoppings of the effective
one-band model are rather insensitive to doping and similar in magnitude.
For large values of $U_{pd}$ the last two terms in Eq. (\ref{inter}) can not
be neglected and affect somewhat the magnitude of these hoppings. As for the
case of the correlated hoppings, if $U_{pd}\lesssim 1$ the different
effective nearest-neighbor repulsions are similar between them and rather
independent of hopping. However in contrast to the former, the $V`s$
increase significantly with $U_{pd}$. For small $\Delta $ and large $U_{pd}$%
, in spite of the fact that there is a significant charge transfer from Cu
to O with doping, the effective repulsions do not change very much because
the decrease (increase) in Cu-O repulsion is approximately compensated by
the increase (decrease) in O-O repulsion.

In Figs. 5 and 6 we show the dependence of the expectation value $%
n_d=\langle n_{i\uparrow }$ $+n_{i\downarrow }\rangle $ on $\Delta $. As
expected, $n_d$ increases with $\Delta $. For small $U_{pd}$ the results
obtained by exact diagonalization (ED) and slave bosons (SB) are
qualitatively similar, although there are quantitative differences of the
order of 20\% for larger dopings. For larger values of $U_{pd}$ there are
some important qualitative differences between both methods which are
discussed below, however both methods coincide in that for intermediate and
small values of $\Delta $, $n_d$ {\it decreases} with doping as a
consequence of Cu-O charge transfer which overcompensates the effect of
doping (see Figs. 6 and 7). This is related with the increase in Cu-O
covalency and the {\it metallization} mentioned before. In contrast, for
large values of $\Delta ,$ $n_d$ {\it increases} faster with doping than for
small values of $U_{pd}$. As a consequence of these opposite effects of 
$U_{pd}$ for small and large $\Delta $, the dependence of $n_d$ with $\Delta $
becomes more abrupt for large $U_{pd},$ eventually driving the valence
instability (VI). The effect of reducing $U_d$ induces a larger Cu occupancy 
$n_d$ at large values of $\Delta $ favoring the VI. The static valence
susceptibility is defined by:

\begin{equation}
\chi _V=\frac{\partial (n_d-n_p)}{\partial \Delta }|_n=2\frac{\partial n_d%
}{\partial \Delta }|_n,  \label{xi}
\end{equation}

\noindent where $n$ is the total occupation per cell and $n_p=n-n_d$ is the O
occupation per cell. $\chi _V$ differs from the charge-transfer
susceptibility $\chi _{CT}$ because the latter is defined at constant
chemical potential instead of at constant $n$. The valence susceptibility $%
\chi _V$ is shown in Fig. 8. For each value of $x=n-1$, there is a
critical value of $U_{pd}$ ($U_{pd}^c(x)$) for which $\chi _V$ diverges
at a critical value of $\Delta $ ($\Delta ^c(x)$), indicating the presence
of a VI. For $U_{pd}>U_{pd}^c$ there is a discontinuous transition as shown
for example in Fig. 6 for $x=0.8$. The minimum possible value of $%
U_{pd}^c(x) $ occurs for maximum doping ($x=1$). The critical value $%
U_{pd}^c(x)$ increases with decreasing $x$. Note that $U_{pd}^c(x)$ found
with SB is always lower than that found with ED. Also, at $x=0$ the SB
approximation gives an artificially large $\chi _V$ at the
metal-insulator transition due to the vanishing of the double occupancy ($d$%
) in the insulating phase and the exaggerated large increase of $d$ 
with $\Delta$ in the
metallic phase near the metal-insulator boundary. Within the ED, $d$ grows
smoothly with $\Delta$ in absence of the VI as one expects.

In Fig. 9 we show the phase-separation diagram calculated with the SB
approximation. Contrary to previous strong-coupling mean-field
approximations, where always one of the phases between which phase
separation takes place has $x=0$ \cite{raimondi}, we obtain, in qualitative
agreement with weak-coupling approximations \cite{kothe} that phase
separation takes place between a phase with large doping $x=1$ and another
one with $x\,$depending of the parameters. As expected, the region of phase
separation grows with increasing $U_{pd}$. At least for $U_{pd}\leq 1,$( 
{\it i.e. }for parameters near those obtained by
constrained-density-functional approximation \cite{lda}) there is no phase
separation. The effect of decreasing $U_d$ on phase separation is to
suppress it for large values of $\Delta $. This is related with the change
of the character of the local singlet $|i2\rangle $ (Eq. \ref{para1}) from
mainly Cu-O or O-O to a doubly occupied Cu orbital. For low values of $%
\Delta $, the phase diagram is rather independent of $U_d.$ Also, the
effects of $U_{pd}$ and $U_d$ on the phase diagram on a qualitative level
can be understood on general physical grounds: increasing these repulsions
favors localization, inhibits the kinetic energy terms, and as a result the
dependence of the energy on occupation is more flat, favoring phase
separation. Although using ED we have only a few possible different
densities, due to the small size of the cluster, the results for the
compressibility at these densities agree qualitatively with the SB\ phase
separation diagram. However there are some important differences in a
quantitative level: a) there is no phase separation for $U_{pd}\leq 2$. b)
For $U_{pd}=3$ there is a small island inside which the system phase
separates in two phases, both with compositions inside the interval $0<x<1$.
Some of these features can be seen in Fig. 10, where we show the chemical
potential $\mu $ as a function of doping. In table I we list the densities
for which the compressibility is negative (a sufficient but not necessary
condition for phase separation) according to our ED results for different
parameters.

\section{Summary and Discussion}

In this work, we have studied the effects of Cu-O Coulomb repulsion $U_{pd}$
on valence instabilities and phase separation. Using a variational form of
the cell method, we derive a one-band effective model with
doping dependent parameters. This model has the form of a generalized
Hubbard with correlated hoppings and nearest-neighbor repulsions. This
approach allows us to study realistic values of the on-site Cu repulsion $%
U_d\sim 10$ eV  and is accurate enough for moderate values of $U_{pd}$ and
on-site O repulsion $U_p$. We obtain that for parameters near those
derived using constrained-density-functional approximation \cite{lda}, in
particular $U_{pd}\lesssim 1.3$ eV, the low- energy physics of the three-band
Hubbard model can be well represented by the effective one-band   model,
with doping independent parameters. The effect of $U_{pd}$ is merely to
renormalize the effective parameters and there are neither valence 
instabilities nor phase separation. 

For larger $U_{pd}$ the effective parameters become strongly doping
dependent and for $U_{pd}\sim 3$ eV valence instabilities and phase
separation appear. In agreement with previous weak-coupling results \cite
{kothe}, we obtain that phase separation starts to occur for large doping
values ($x\sim 0.7\pm 0.1$ according to the exact-diagonalization results
or $x\sim 1$ according to the slave-boson ones). The valence instability
begins at $x=1$ in both treatments. This facts are in contrast to results of
alternative previous strong-coupling approaches \cite{hicks,raimondi,hira}.
This is probably due to the facts that on the one hand, the slave-boson
approximation produces an artificial increase of $\chi _V$ at the
metal-insulator transition ($x=0$), as discussed in section III, favoring PS
at low $x$. On the other hand in Refs. \cite{hicks,raimondi,capra}, the $%
U_{pd}$ and $U_p$ terms were taken in a mean-field approximation,
essentially equivalent to Hartree-Fock , (and since $U_d$ is large, this
induces an artificially large increase of the energy with doping), while in
our present treatment most of the $U_{pd}$ term and a large part of the $U_p$
one are treated exactly. In fact, we have verified that treating both terms
in Hartree-Fock before performing the low-energy reduction to the one-band
model (as was done in previous studies of the metal-insulator transition 
\cite{sim2}), the results become qualitatively similar to those of Refs. 
\cite{hicks,raimondi}.

\section{Acknowledgments.}

We are indebted to E. R. Gagliano for providing us the program for the exact
diagonalization. One of us (M.E.S.) would like to thank M. Bali\~{n}a for
very  helpful discussions. M. E. S. is supported by the Consejo Nacional de
Investigaciones Cient\'{\i}ficas y T\'{e}cnicas (CONICET), Argentina. A. A.
A. is partially supported by CONICET.

\newpage

\section*{Figure Captions}

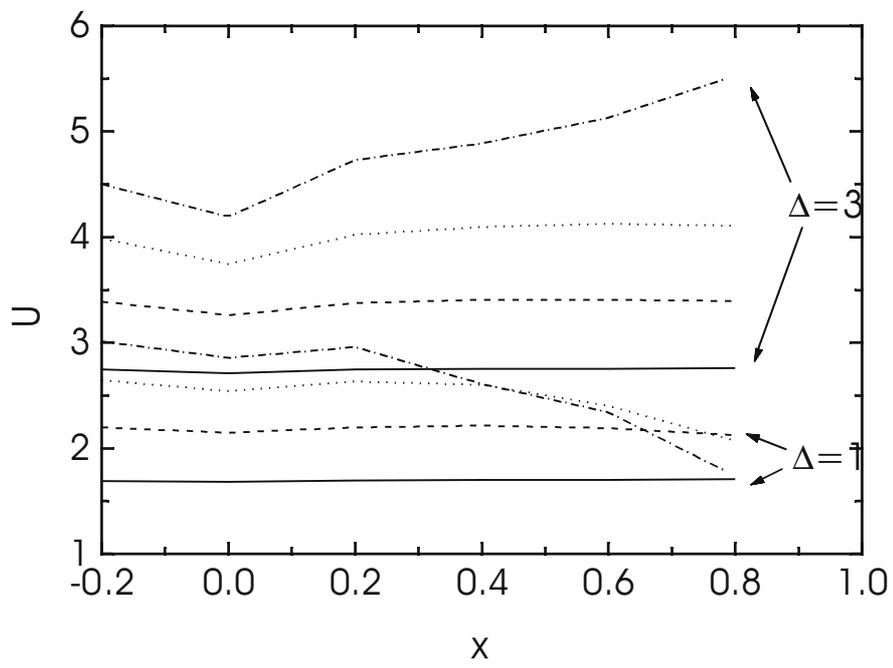
\noindent{\bf Fig. 1:} On-site repulsion of the effective one-band model
(Eq. \ref{1b}) obtained by exact diagonalization (ED) (see text) as a
function of hole doping $x$ for different values of the Cu-O charge transfer
energy ($\Delta $) and Cu-O repulsion: $U_{pd}$=1 (solid), 2 (dashed), 3
(dotted), 4 (dotted-dashed). Other parameters of the original three-band
model (Eq. \ref{3b} ) are: $U_d=7$, $U_p=4$, $t_{pd}=1$, $t_{pp}=.5$.

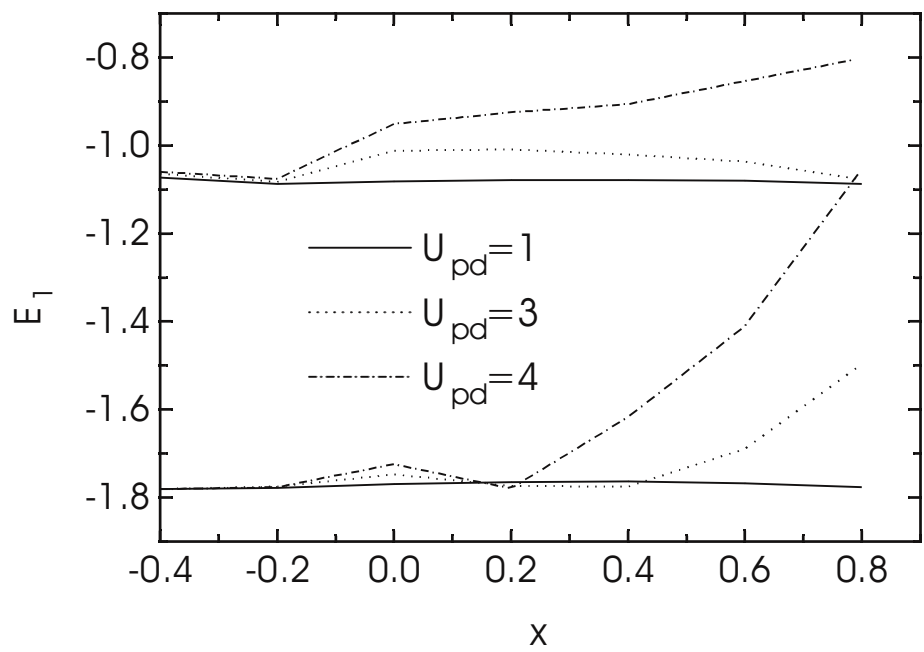
\noindent{\bf Fig. 2:} One particle on-site energy of the effective one-band
model (Eq. \ref{1b}) as a function of doping for different values of $\Delta 
$ and $U_{pd}.$ The three lower curves correspond to $\Delta =1$ and the
other three to $\Delta =3$. Other parameters as in Fig. 1.

\noindent{\bf Fig. 3:} Nearest-neighbor hoppings of the effective one-band
model (Eq. \ref{1b}) as a function of doping for different values of $\Delta 
$ ((a) $\Delta =1$, (b) $\Delta =3$) and $U_{pd}$. Other parameters as in
Fig. 1.

\noindent{\bf Fig. 4:} Nearest-neighbor Coulomb repulsions of the effective
one-band model (Eq. \ref{1b}) as a function of doping for different values
of $\Delta $ and $U_{pd}$=1 (solid), and 3 (dotted). Other parameters as in Fig.
1.

\noindent{\bf Fig. 5:} Cu occupancy as a function of Cu-O charge transfer
energy ($\Delta $) for different dopings (indicated at the right of the
corresponding curve) obtained by (a) exact diagonalization  and (b) slave bosons.
Other parameters are $U_{pd}=2$, $U_d=7$, $U_p=4$, $t_{pd}=1$%
, $t_{pp}=0.5.$

\noindent{\bf Fig. 6:} Same as Fig. 5 for $U_{pd}=4,$ $U_d=10$.

\noindent{\bf Fig. 7:} Cu occupancy as a function of doping for different
values of $\Delta $ (indicated at the right of the corresponding curve) and $%
U_d$, obtained by slave bosons (SB). Other parameters are $U_{pd}=3$, $U_p=4$%
, $t_{pd}=1$, $t_{pp}=0.5.$

\noindent{\bf Fig. 8:} Valence susceptibility as a function of $\Delta $ for
the different positive dopings of Fig. 6 obtained by (a) exact diagonalization
 and (b) slave bosons. Other parameters as in Fig. 6.

\noindent{\bf Fig. 9:} Phase-separation diagram $\Delta $ vs. $x$ obtained
using SB\ for different values of $U_d$ and $U_{pd}$. The dotted line
indicates the first-order valence transition for $U_{pd}=4$ (for $U_{pd}=2,3$
there are no first order VI\ transition). The solid squares represent the
same obtained with ED. Other parameters as in Figs. 1,5.

\noindent{\bf Fig. 10:} Chemical potential as a function of doping
for several values of $\Delta $ calculated with (a) SB and (b) ED. Other
parameters are $U_{pd}=3$, $U_d=7$, $U_p=4$, $t_{pd}=1$, $t_{pp}=0.5.$

\section{table caption}

\noindent{\bf Table I: }Values of the density $n$ for which the
compressibility is negative according to ED for different values of $U_d$, $%
U_{pd}$ and $\Delta $. Other parameters are   $U_p=4$, $t_{pd}=1$, $%
t_{pp}=0.5. $

$
\begin{tabular}{||l|l|l|l||l|l|l|l||}
\hline\hline
$U_d$ & $U_{pd}$ & $\Delta $ & $n$ & $U_d$ & $U_{pd}$ & $\Delta $ & $n$ \\ 
\hline\hline
7 & 3 & 0.5 & 0.4-0.8 & 10 & 3 & 0.5 & 0.4-0.8 \\ \hline
7 & 3 & 1 & 0.4-0.8 & 10 & 3 & 2.5 & 0.4-0.8 \\ \hline
7 & 3 & 2 & -- & 10 & 3 & 3 & --- \\ \hline
7 & 4 & 0.5 & 0.2-1 & 10 & 4 & 0.5 & 0-1 \\ \hline
7 & 4 & 1 & 0.4-1 & 10 & 4 & 1 & 0-1 \\ \hline
7 & 4 & 2 & 0.6-1 & 10 & 4 & 2 & 0.4-1 \\ \hline
7 & 4 & 3 & -- & 10 & 4 & 3 & 0.6-1 \\ \hline
&  &  &  & 10 & 4 & 4 & 0.8-1 \\ \hline
&  &  &  & 10 & 4 & 5 & --- \\ \hline\hline
\end{tabular}
$


\begin{references}
\bibitem{eme}  V. J. Emery, Phys. Rev. Lett. {\bf 58}, 2794 (1987).

\bibitem{var}  C. M. Varma, S. Schmitt-Rink, and E. Abrahams, Solid State
Commun. {\bf 62}, 681 (1987).

\bibitem{little}  P. B. Littlewood, C. Varma, and E. Abrahams, Phys. Rev.
Lett. {\bf 63}, 2602 (1989).

\bibitem{zha}  F. C. Zhang and T. M. Rice, Phys. Rev. B {\bf 37}, 3759
(1988).

\bibitem{rieri}  E. Dagotto and J. Riera, Phys. Rev. Lett. {\bf 70}, 682
(1993); E. S. Heeb and T. M. Rice, Europhys. Lett. {\bf 27}, 673 (1994).

\bibitem{rvb}  C. D. Batista and A. A. Aligia, LT5375B.

\bibitem{rush}   Han Rushan, C. Chew, K. Phua, and Z. Gan, Phys. Rev. B {\bf %
39}, 9200 (1989).

\bibitem{lda}  A. K. Mc Mahan, J. Annett, and R. Martin, Phys. Rev B {\bf 42}%
, 6268 (1990); M. Hybertsen, M. Schl\"{u}ter, and N. Christensen, Phys. Rev B%
 {\bf 39}, 9028 (1989); M. Hybertsen, E. Stechel, M. Schl\"{u}ter, and D.
Jennison, Phys. Rev. B {\bf 41}, 11068 (1990).; J. Grant and A. Mc Mahan,
Phys. Rev. Lett. {\bf 66}, 488 (1991).

\bibitem{cluster}  E. R. Gagliano, A. G Rojo, C. A. Balseiro, and B.
Alascio, Solid State Commun. {\bf 64}, 901 (1987); C. A. Balseiro,
A. G. Rojo, E. R. Gagliano, and B. Alascio, Phys. Rev. {\bf B} 38,
9315 (1988); J. E. Hirsch, S. Tang, E Loh Jr., and D. J. Scalapino,%
Phys. Rev. Lett .{\bf 60}, 1668 (1988).

\bibitem{sudbo}  A. Sudb\o , C. M. Varma, T. Giamarchi, E. B. Stechel, and
R. T. Scalettar, Phys. Rev. Lett. {\bf 70}, 978 (1993).

\bibitem{perakis}  I. E. Perakis, C. M. Varma, and A. E. Ruckestein, Phys.
Rev. Lett. {\bf 70}, 3867 (1993); C. M. Varma, Phys. Rev. Lett. {\bf 75 }, 898
(1995).

\bibitem{hicks}  J. Hicks, A. Ruckenstein, and S. Schmitt-Rink, Phys. Rev. B 
{\bf 45}, 8185 (1992)

\bibitem{bang}  Y. Bang, G. Kotliar, R. Raimondi, C. Castellani, and M.
Grilli, Phys. Rev. B {\bf  47}, 3323 (1993).

\bibitem{raimondi}  R. Raimondi, C. Castellani, M. Grilli, Y. Bang, and G.
Kotliar, Phys. Rev. B {\bf  47}, 3331 (1993); M. Grilli, R. Raimondi, C.
Castellani, C. Di Castro, and G. Kotliar, Phys. Rev. Lett. {\bf 67}, 259
(1991).

\bibitem{little2}  P. B. Littlewood, Phys. Rev. B {\bf 42},10075 (1990).

\bibitem{kothe}  N. Kothekar, K. Quader, and D. Allender, Phys. Rev. B {\bf %
51}, 5899 (1995).

\bibitem{copper}  S. N. Coppersmith and P. B. Littlewood, Phys. Rev. B {\bf %
41}, 2646 (1990).

\bibitem{hira}  D. Hirashima, Y. Ono, T. Matsuura, and Y. Kuroda, J. Phys.
Soc. Jpn. {\bf 61}, 649 (1992).

\bibitem{capra}  S. Caprara and M. Grilli, Phys. Rev. B{\bf \ 49}, 10805
(1994).

\bibitem{fed}  H. B. Sch\"{u}ttler and A. J. Fedro, Phys. Rev. B {\bf 45},
7588 (1992).

\bibitem{jef}  J. H. Jefferson, H. Eskes, and L. F. Feiner, Phys. Rev. B 
{\bf 45}, 7959 (1992)

\bibitem{fei}  L. F. Feiner, J. H. Jefferson, and R. Raimondi, preprint.

\bibitem{hay}  R. Hayn, V. Yushankhai, and S. Lovstov, Phys. Rev. B {\bf 47}%
, 5253 (1993), and references therein.

\bibitem{sim1}  M. E. Sim\'{o}n, M. Bali\~{n}a, and A. A. Aligia, Physica C 
{\bf 206}, 297 (1993).

\bibitem{sim2}  M. E. Sim\'{o}n and A. A. Aligia, Phys. Rev. B {\bf 48},
7471 (1993).

\bibitem{bel}  V. I. Belinicher and A. L. Chernyshev, Phys. Rev. B {\bf 49},
9746 (1994); V. Belinicher, A. Chernyshev, and L. Popovich, Phys. Rev. B 
{\bf 50}, 13768 (1994).

\bibitem{ali}  A. A. Aligia, M. E. Sim\'{o}n, and C. D Batista, Phys. Rev. B 
{\bf 49}, 13061 (1994).

\bibitem{bati}  C. D. Batista and A. A. Aligia, Phys. Rev. B {\bf 47}, 8929
(1993).

\bibitem{tripletes}  M. E. Sim\'{o}n and A. A. Aligia, Phys. Rev B {\bf 52},
7701 (1995).

\bibitem{api}  R. Raimondi, L. Feiner, and J. Jefferson, Physica C {\bf 235-240}, 2203 (1994).

\bibitem{bati1}  C. D. Batista and A. A. Aligia, Phys. Rev. B {\bf 48}, 4212
(1993); E {\bf 49}, 6436 (1994).

\bibitem{ero}  J. M. Eroles, C. D. Batista, and A. A. Aligia, BD5722.

\bibitem{kot}  G. Kotliar and A. E. Ruckenstein, Phys. Rev. Lett. {\bf 57},
1362 (1986).

\bibitem{hall}  N. P. Ong, Z. Z. Wang, J. Clayhold, J. M. Tarascon, L. H.
Greene, and W. R. McKinnon, Phys. Rev. B {\bf 35}, 8807 (1987); K. Sreedhar
and P. Ganguly, Phys. Rev. B {\bf 41}, 371 (1990)
\end{references}
\end{document}